\documentclass[letterpaper, 10 pt, conference]{ieeeconf}
\IEEEoverridecommandlockouts

\usepackage{algorithmic}
\usepackage{algorithm}
\usepackage{array}
\usepackage{textcomp}
\usepackage{stfloats}
\usepackage{url}
\usepackage{verbatim}
\usepackage{graphicx}
\usepackage{cite}
\usepackage[colorlinks=true, linkcolor=blue, citecolor=blue, urlcolor=blue]{hyperref}

\usepackage{amsmath,amssymb,amsfonts}
\usepackage{algorithmic}
\usepackage{graphicx,epstopdf}
\usepackage{textcomp}
\usepackage{picinpar}
\usepackage{url}
\usepackage{flushend}
\usepackage{colortbl}
\usepackage{soul}
\usepackage{multirow}
\usepackage{pifont}
\usepackage{xcolor}
\usepackage{alltt}
\usepackage{enumerate}
\usepackage{siunitx}
\usepackage{url}
\usepackage{pbox}
\usepackage{bbding}
\usepackage{amssymb}
\usepackage{tikz}
\usepackage{lipsum} % 
\usepackage{multicol} % 
\usepackage{xcolor}
\usepackage{hyperref}
\usepackage{tabularx} 
\usepackage{booktabs}
\usepackage{fix-cm}

\usepackage{amsthm}
\usepackage{bm} 
\usepackage{mathrsfs}

\usepackage{array}
\usepackage{empheq} %%%%%%%%%%%%%%%%%%%%%%% SE LO PUSE
\usepackage{cases}

\usepackage{caption}
\usepackage{graphicx}
\usepackage{subcaption}

\definecolor{BlueViolet}{rgb}{0.36, 0.4, 0.8}

\makeatletter
\def\IEEElabelanchoreqn#1{\bgroup
\def\@currentlabel{\p@equation\theequation}\relax
\def\@currentHref{\@IEEEtheHrefequation}\label{#1}\relax
\Hy@raisedlink{\hyper@anchorstart{\@currentHref}}\relax
\Hy@raisedlink{\hyper@anchorend}\egroup}
\makeatother

\title{The QuadSoft: Design, Construction, and Experimental Validation of a Soft and Actuated Quadrotor}

\author{Rodolfo Verd\'in\textsuperscript{1}, Hugo Moreno\textsuperscript{1}, Mark W. Spong\textsuperscript{2}, and Gerardo Flores\textsuperscript{3}
       \thanks{\textsuperscript{1}Laboratorio de Percepción y Robótica, Centro de Investigaciones en Óptica, León 37150, México. Emails: \texttt{rverdin@cio.mx}, \texttt{hugoamj@cio.mx}}
        \thanks{\textsuperscript{2}Erik Jonsson School of Engineering \& Computer Science, Department of Systems Engineering, University of Texas at Dallas, Richardson, TX 75080, USA. Email: \texttt{mspong@utdallas.edu }}
        \thanks{\textsuperscript{3}RAPTOR Lab, School of Engineering, College of Arts and Sciences, Texas A\&M International University, Laredo, TX 78041 USA. (Corresponding author Email: \texttt{gerardo.flores@tamiu.edu})}
}

\begin{document}

\maketitle
\thispagestyle{empty}
\pagestyle{empty}

\begin{abstract}
This paper presents \textit{QuadSoft}, a novel fully actuated quadrotor equipped with continuous-curvature, tendon-driven soft robotic arms. The design combines a semi-rigid central frame with flexible arms, enabling controlled structural reconfiguration during flight without altering the propeller layout. Unlike existing soft aerial platforms that rely on discrete bending joints, QuadSoft utilizes a continuum deformation approach to modulate arm curvature, actively adjusting its thrust vector and aerodynamic characteristics. We characterize the geometric mapping between servomotor input and the resulting constant curvature, validating it experimentally. Outdoor flight tests demonstrate stable take-off, hover, directional maneuvers, and landing, confirming that controlled arm bending can generate horizontal displacement while preserving altitude. Measurements of pitch, roll, and curvature angles show that the platform follows intended actuation patterns with minimal attitude deviations. These results demonstrate that QuadSoft preserves the baseline stability of rigid quadrotors while enabling morphology-driven maneuverability, all under the standard PX4 autopilot without retuning. Beyond a proof of concept, this work establishes a distinctive outdoor validation of a tendon-driven continuum morphing quadrotor, opening a new research avenue toward adaptive aerial systems that combine the safety and versatility of soft robotics with the performance of conventional UAVs.
\end{abstract}
%
%
%\begin{keywords}
%Soft robotics, UAVs, Quadrotors, Mechanical design, Morphological adaptation, Fully actuated, Experimental validation
%\end{keywords}
%
%
\section*{Supplementary material}
A video showing the experimental results is also available at the following link: \textcolor{blue}{\url{https://www.youtube.com/watch?v=txPaA6l5GvE}}.
\section{Introduction}
Quadrotors are widely used in aerial robotics for their simplicity and agility, yet rigid-frame designs limit adaptability in cluttered or dynamic environments. To address this, we propose \textit{QuadSoft}, a novel platform with tendon-driven flexible arms that can \textbf{reconfigure in flight} through \textbf{continuous-curvature deformation}. Unlike conventional morphing designs that rely on rigid hinges or discrete joints, QuadSoft leverages soft morphology to achieve thrust vectoring while maintaining structural integrity. This capability enables navigation in tight spaces, improved aerodynamic efficiency, and enhanced stability. By combining a continuum mechanics approach with standard flight control, QuadSoft bridges the gap between compliant robotics and the need for versatile, resilient aerial platforms capable of operating in real-world conditions.
%
%%%%%%%%%%%%%%%%%%% FIGURE %%%%%%%%%%%%%%%%%%
\begin{figure}[t]
  \hfill
  \includegraphics[width=\columnwidth]{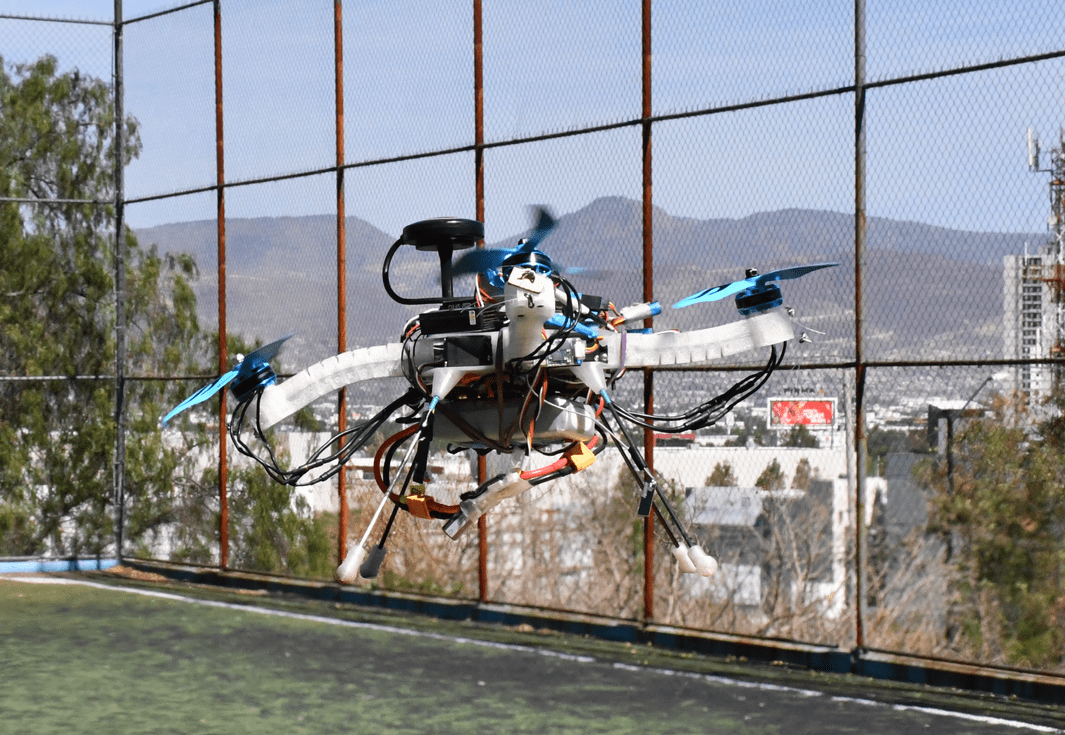} % 
  \caption{QuadSoft prototype during outdoor flight. The four arms are tendon-driven and flexible, actuated by servomotors to enable in-flight morphological adaptation.}
  \label{fig:model}
\end{figure}
%%%%%%%%%%%%%%%%%%% FIGURE %%%%%%%%%%%%%%%%%%
%
%%%%%%%% STATE OF ART %%%%%%%%%%%%%%%%%%%%%%%
\subsection{State of the Art}\label{sec:state-of-art}
Quadrotors have become central in aerial robotics for their efficiency, stability, and simplicity, yet their rigid-frame architectures inherently limit adaptability in cluttered or dynamic environments \cite{Xu, Bucki, Falanga, Sihite2023}. Soft robotics has shown that flexible, biologically inspired structures can enhance versatility and safety across manipulators and ground robots \cite{KIM2013287, Rus2015, Jeger2024, rus2025controlling, MORENO2025103407, guan2023trimmed, stella2025synergy, trompetto2022soft, kanzler2022soft}, with pneumatic, tendon-driven, and smart-material actuators enabling precise shape control and compliance \cite{jung2024untethered, li2023pedot, zhang2025thinfilm, kim2025fiber, li2023deepsea}. However, the application of these principles to aerial vehicles remains limited \cite{Nguyen, Hwang, Haluska}, with only partial solutions addressing flexibility through distributed-parameter modeling and delay-resistant adaptive control \cite{doi:10.2514/1.G007665}.

Existing morphing quadrotors each fall short in different ways. Impact-resilient designs like Morphy \cite{Petris} prioritize crash recovery over active reconfiguration. Dual-axis tilting-rotor mechanisms \cite{10752979} achieve 6-DOF control but at the cost of mechanical complexity and additional failure points. Hybrid platforms such as SMORS \cite{Ryll} combine rigid arms with soft components yet offer only partial deformation. Tendon-driven platforms like those presented in \cite{9851515} and \cite{11223764} employ discrete hinge-like bending points and typically require hexarotor configurations to compensate for their limited DOF. In these systems, the tendon actuation is primarily intended for surface grasping rather than for improving maneuverability or flight control. Beyond quadrotors, morphing-wing UAVs \cite{Zhu2023, Tariq} and inflatable structures like SFAR \cite{Jia} target aerodynamic performance and safety, respectively, rather than active in-flight morphological reconfiguration.

In contrast, QuadSoft adopts a constant-curvature continuum approach in a standard X-quadrotor, achieving full 6-DOF actuation with four rotors—without rigid tilting joints, extra propulsion units, or specialized flight control hardware. Notably, the design is validated on a standard, unmodified PX4 autopilot \cite{px4,px41}, as detailed in the following section.
%
%
%%%%%%%%%%%%%%%% CONTRIBUTION %%%%%%%%%%%%%%%
\subsection{Contribution}\label{sec:contributions}
Building on earlier theoretical work on constant-curvature soft aerial vehicles \cite{FLORES2025}, this paper focuses on the design, construction, and real-world experimental validation of QuadSoft. The main contributions are:
\begin{itemize}
\item \textit{Continuum Design and Construction:} A tendon-driven soft arm based on a constant-curvature approach, enabling smooth in-flight reconfiguration within a standard X-quadrotor configuration—without discrete joints or rigid hinges.
\item \textit{Geometric Mapping:} A closed-form mapping from servomotor input to arm curvature is derived and experimentally validated, linking tendon actuation to 6-DOF thrust-vectoring capability.

\item \textit{Outdoor Flight Validation:} First outdoor flight tests of a continuum-morphing quadrotor, demonstrating stable hover and morphology-driven translation using a standard, unmodified PX4 autopilot.

\item \textit{Baseline for Control Allocation:} Physical stability validated up to 22$^\circ$ of curvature (within a 35$^\circ$ design envelope), providing a foundation for future custom mixer development in fully actuated soft UAVs.
\end{itemize}
%
%
%%%%%%%%%%%%% ORGANIZATION %%%%%%%%%%%%%%%%%%%%%%
\subsection{Organization}
The remainder of this paper is organized as follows. Section~\ref{ssec:problem} formulates the central challenge of designing the \textit{QuadSoft} with tendon-driven soft arms. The geometric mapping from servomotor input to arm curvature and propeller orientation is derived in Section~\ref{sec:geometric}. Section~\ref{ssec:Mechelec} describes the mechanical and electronic implementation of the proposed design. Section~\ref{sec:MainResults} presents the outdoor experimental validation, focusing on hover stability and morphology-induced translation. Finally, Section~\ref{conclusion} concludes the paper and outlines future research directions.

%%%%%%% PROBLEM STATEMENT %%%%%%
\section{Problem Statement}\label{ssec:problem}
Designing a morphing quadrotor with soft actuated arms introduces a central challenge: enabling in-flight reconfiguration without compromising flight stability. The inherent flexibility of soft structures allows morphological adaptation but also induces deformations and vibrations that can degrade control performance. Furthermore, maintaining a lightweight structure is critical, as additional mass increases the required lift forces, which in turn amplify structural deformations. This challenge is divided into two interconnected subproblems:  

\begin{itemize}
    \item \textbf{Subproblem A --- Mechanical Design:} Developing a lightweight, tendon-driven arm mechanism that balances compliance with structural stability to minimize oscillations while allowing sufficient curvature for reconfiguration.  
    \item \textbf{Subproblem B --- Integration and Modeling:} Deriving the actuator–curvature–propeller mapping to ensure that morphological changes translate into predictable and controllable motions while preserving baseline stability.  
\end{itemize}

The solution to Subproblem A is detailed in Section~\ref{ssec:Mechelec}, while Subproblem B is addressed in Section~\ref{sec:geometric} and subsequently validated through the outdoor experiments presented in Section~\ref{sec:MainResults}.
\section{Geometric Mapping of Servomotor to Arm Curvature}\label{sec:geometric}
\begin{figure}[ht]
  \centering
  \hfill
\includegraphics[width=\columnwidth]{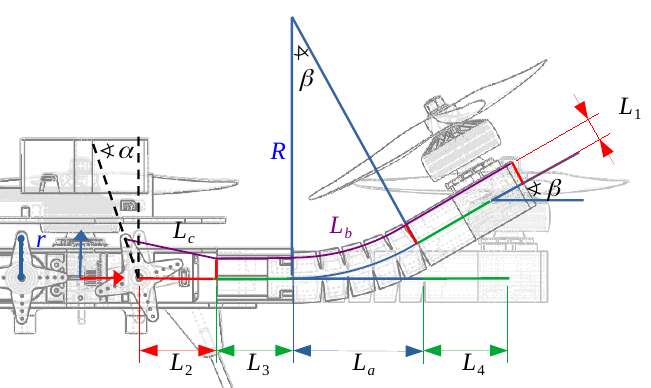}
  \caption{Schematic of the QuadSoft's flexible arm, showing the geometric variables used to calculate the bending angle $\beta$ as a function of the servo motor's rotation angle $\alpha$. The lengths $L_a$, $L_b$, $L_c$, and $L_3$ represent segments of the arm and its base, while $\beta$ describes the arm's curvature due to flexibility.}
  \label{fig:eq}
\end{figure}

The angular position of the propellers with respect to the main body frame can be expressed as a function of the servomotor angle. This relationship arises from the geometric constraints imposed by the tendon-driven arm. Fig.~\ref{fig:eq} illustrates the relevant variables: $L_a$, the fixed arc length determined by the semi-rigid core, and $L_b$, the variable length defined by the tendon pulled by the servomotor. Since the total tendon length remains constant (highlighted in purple in Fig.~\ref{fig:eq}), the change in length $\Delta L_c$ caused by a rotation $\alpha$ of the servomotor can be obtained using trigonometry:  
\begin{align}\label{eq1}
\Delta L_c &= K \Bigg( \sqrt{\big(L_2 + r \sin(\alpha)\big)^2 + \big(r \cos(\alpha) - L_1\big)^2} \notag \\
&\quad - \sqrt{L_2^2 + (r - L_1)^2} \Bigg),
\end{align}
where $r$ is the radius of the tendon trajectory, $\alpha$ is the servomotor angle, and $K$ is a proportionality constant that accounts for unmodeled effects.  

The arc lengths $L_a$ and $L_b$ are related to the arm’s angular displacement $\beta$:  
\begin{equation}\label{eq2}
L_a = \frac{2\pi R \beta}{360}, \quad 
L_b = \frac{2\pi (R - L_1)\beta}{360},
\end{equation}
where $R$ is the nominal radius of the rigid arc, $R - L_1$ corresponds to the effective radius when the tendon is pulled, and $\beta$ is the arm curvature angle. Since $\beta$ is identical in both cases,  
\begin{equation}
\frac{L_a}{R} = \frac{L_b}{R - L_1}.   
\end{equation}
From this relation, the effective radius $R$ is expressed as a function of $L_a$, $L_b$, and $L_1$:  
\begin{equation}\label{eq4}
R = \frac{L_a L_1}{L_a - L_b} = \frac{L_a L_1}{\Delta L_c}.
\end{equation}
Finally, the curvature angle $\beta$ is as follows:
\begin{equation}\label{eq5}
\beta = \frac{L_a \cdot 360}{2\pi R}.
\end{equation}

Together, these expressions provide a complete geometric mapping from servomotor input $\alpha$ to arm curvature $\beta$, and thus to the propellers’ angular orientation relative to the body frame.  

%
%%%%%%%%%%% f i g u r e %%%%%%%%%%%%%%
%%%%%%%%%%% f i g u r e %%%%%%%%%%%%%%
%

Experimental measurements further showed that $\beta$ can be well approximated by a cubic interpolation of the form:  
\begin{equation}\label{eq6}
\beta_i(\alpha) = a_i (\alpha - \alpha_i)^3 + b_i (\alpha - \alpha_i)^2 + c_i (\alpha - \alpha_i) + d_i,
\end{equation}
where the coefficients $(a_i, b_i, c_i, d_i)$ are obtained via data fitting for each arm.  

\begin{figure*}[t]
  \centering  
\includegraphics[width=\textwidth]{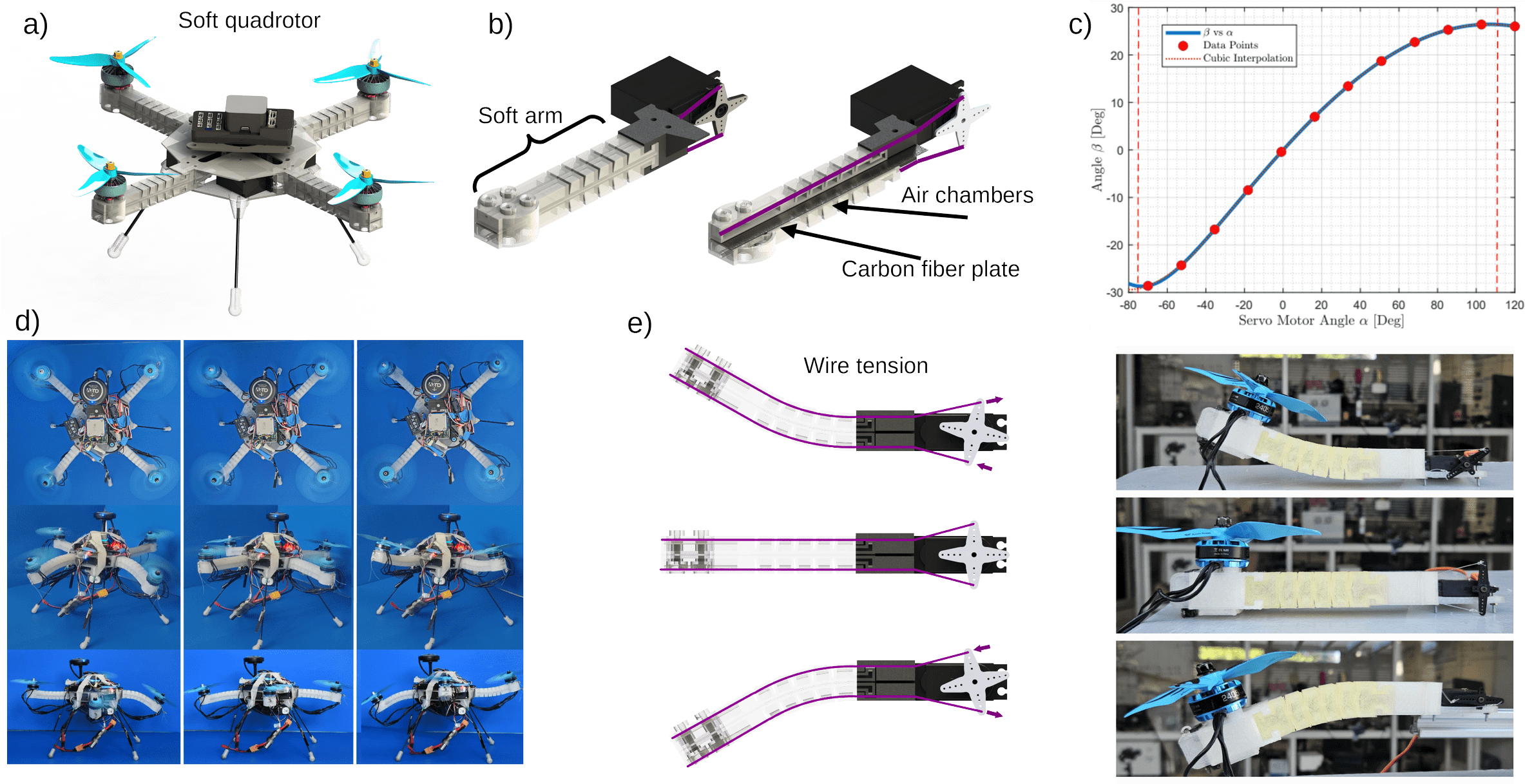}
  \caption{Design and actuation of the QuadSoft. 
(a) Complete QuadSoft platform. 
(b) Soft arm with semi-rigid carbon fiber insert and air chambers for vibration damping. 
(c) In the graph, the geometric mapping between the servo input $\alpha$ and the arm curvature $\beta$ is shown in blue, while the cubic interpolation obtained using Eq.\ref{eq6} is shown as a red dashed line.
(d) Photographs of the QuadSoft with arms at different bending angles. 
(e) Dual-cable tendon mechanism: one cable induces positive curvature, the other negative. 
This design ensures precise and reliable arm reconfiguration while balancing flexibility and structural stability.}
  \label{fig:mechanism}
\end{figure*}

Fig.~\ref{fig:mechanism}(c) illustrates this mapping, showing the experimental $\alpha$–$\beta$ curve along with the operational limits beyond which vehicle lift would be compromised \cite{AFLORES2023}.
%
%
%%%%%%%%%%%%%%%% FIGURE %%%%%%%%%%%%%%%%%%%%%
%%%%%%%%%%%%%%%% FIGURE %%%%%%%%%%%%%%%%%%%%%
%
%%%%%%%%%%%%%%%%%%%%%%%%%%%%
\begin{figure}[t]
   \centering  
\includegraphics[width=\columnwidth]{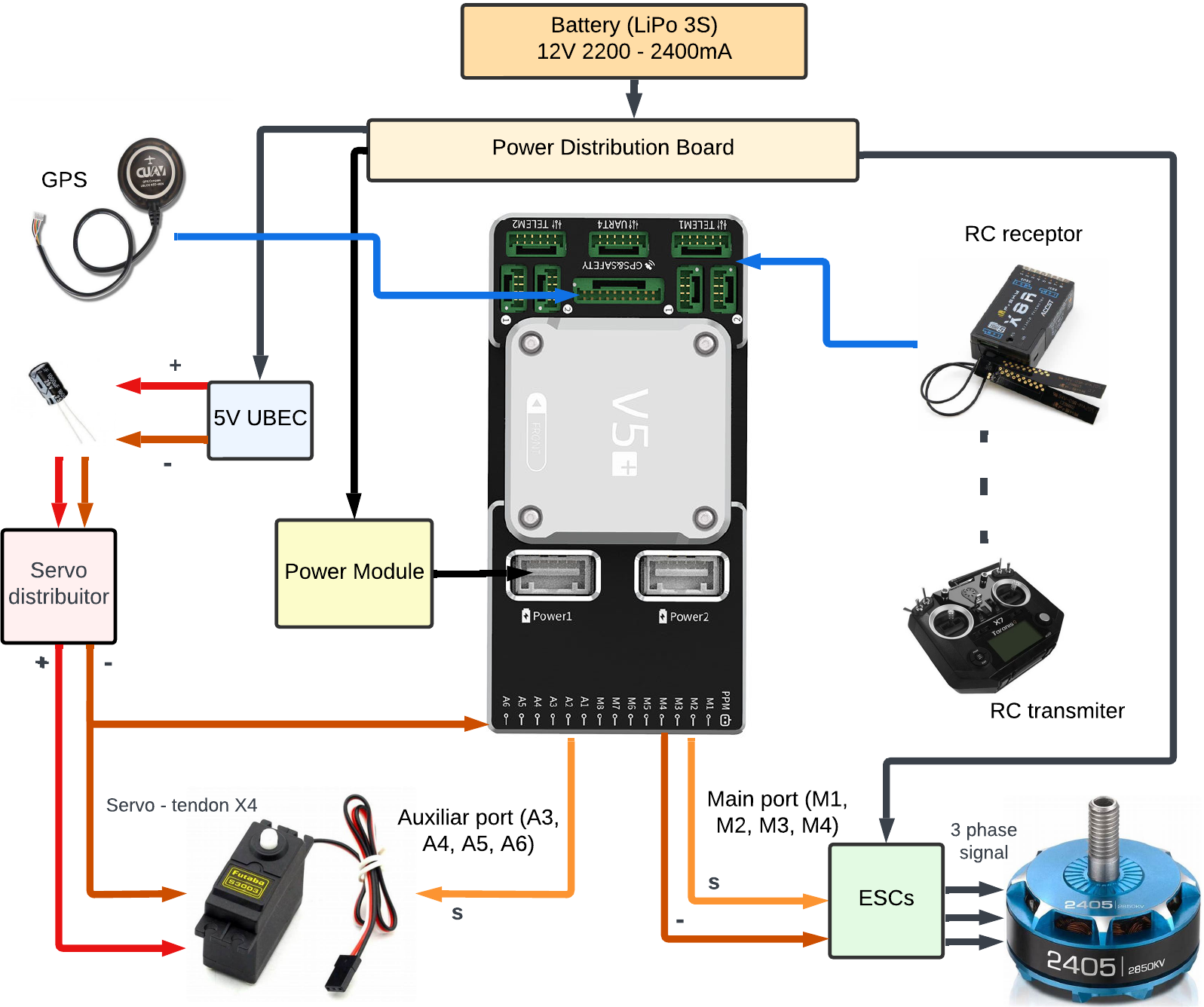}
  \caption{Electrical system diagram of the drone with a PX4 controller. It shows power distribution from a LiPo battery, motor control via ESCs, and servo connections through a distributor and UBEC. A capacitor reduces electrical noise, and the servos adjust tendon angles for morphing. The RC receiver enables remote control.
  }
  \label{fig:electronic}
\end{figure}
%%%%%%%%%%%%%%%%%%% figure %%%%%%%%%%%%%%%%%%%%%%%%%%%%
%
%

%
%%%%%%%%%%%% S E C T I O N %%%%%%%%%%%
\section{Mechanical and Electronic Design}\label{ssec:Mechelec}

\subsection{Structural Design and Stiffness}\label{ssec:StructuralDesign}
The soft arm features a hybrid structure—a flat carbon fiber core embedded in a TPU matrix with internal air cavities—designed for \textbf{anisotropic stiffness}. This planar core provides high compliance for vertical bending while remaining exceptionally stiff in torsion, preventing unwanted yaw-axis twisting. Furthermore, the TPU and air chambers act as a \textbf{vibration damper}, dampening high-frequency motor vibrations to preserve baseline flight stability. The platform's mass distribution is detailed in Table~\ref{tab:mass_budget}.

\subsection{Actuation Mechanism}\label{ssec:ActuationMech}
Active reconfiguration relies on two antagonistic nylon tendons connecting a base servomotor to a tip pulley (Fig.~\ref{fig:mechanism}b). Adjusting tendon tension induces continuous positive or negative bending (Fig.~\ref{fig:mechanism}e). The arm design has a physical bending limit of 28$^\circ$, since propeller efficiency decreases beyond 25$^\circ$. This limit is set by the gap between the arm segments, which prevents exceeding the allowable curvature and avoids propeller-arm collisions. By physically constraining the maximum bending angle, the design also mitigates material fatigue. Moreover, adjusting the segment spacing during manufacturing provides a simple parametric means to tune arm compliance. Arm specifications are listed in Table~\ref{tab:flexible_arms}.

%\textcolor{red}{The internal TPU cavities strategically act as passive mechanical limiters, physically constraining the maximum bending angle to prevent material fatigue. Adjusting their spacing during manufacturing provides a simple parametric method to tune arm compliance. Arm specifications are listed in Table~\ref{tab:flexible_arms}.}
%
%
%%%%%%%%%%%%%%%%%%%% table %%%%%%%%%%%%%%%%%%
\begin{table}[ht]
\centering
\renewcommand{\arraystretch}{1.2}
\begin{tabularx}{\columnwidth}{X X X X}
\hline
\textbf{Characteristic} & \textbf{Value} & \textbf{Characteristic} & \textbf{Value} \\
\hline
Material                 & TPU & Length (mm)              & 109.5            \\
Width (mm)               & 20.0          & Height (mm)              & 17.5             \\
Mass (g)                & 32.0          & Elastic modulus (MPa)    & 30.0             \\
Tensile strength (MPa)  & 40.0          & Flexural modulus (MPa)   & 26.0             \\
Poisson's ratio         & 0.48          & Tendon type              & Nylon string      \\
Tendon diameter (mm)    & 0.8           &                          &                  \\
\hline
\end{tabularx}
\caption{Specifications of the flexible arms and tendons used in the morphing mechanism of the QuadSoft. These properties ensure lightweight construction, mechanical flexibility, and structural integrity for stable flight.}
\label{tab:flexible_arms}
\end{table}
%
%%%%%%%%%%%table%%%%%%%%%%%%%%%%%%%%%%%%%%%%%%%
\begin{table}[ht]
\centering
\renewcommand{\arraystretch}{1.2}
\begin{tabular}{l c c}
\hline
\textbf{Unit} & \textbf{Weight (g)} & \textbf{ Weight Percentage (\%)} \\
\hline
4 × (Motors and propellers) & 126 & 11 \\
3S LiPo battery & 250 & 22 \\
Chassis & 81 & 7.1 \\
4 × (servo motors) & 28 & 2.4 \\
Flight controller & 50 & 4.4 \\
4 × (Deformable tendons) & 32 & 6.7 \\
Other electronics & 277 & 24.5 \\
Total Weight & \textbf{980 g} & \textbf{100}\\
\hline
\end{tabular}
\caption{Mass budget of the proposed soft-Quadrotor, detailing the absolute and relative weight contributions of each component.}
\label{tab:mass_budget}
\end{table}
%%%%%%%%%%%table%%%%%%%%%%%%%%%%%%%%%%%%%%%%%%
%
%
\subsection{Electronics Integration}\label{ssec:electronic}
The electronic architecture is deliberately designed to demonstrate that complex soft-morphing can be managed by standard, off-the-shelf avionics. The core of the system is a Pixhawk\textsuperscript{\textregistered} V5+ autopilot running an unmodified PX4 flight stack. 

Power is supplied by a 12V 3S LiPo battery via a Power Distribution Board (PDB). The Pixhawk simultaneously manages baseline flight stabilization—commanding the brushless motors via standard ESCs—while synchronizing the servomotors responsible for tendon actuation. To prevent voltage drops during high-torque reconfigurations, a dedicated UBEC regulates a consistent 5V supply to the servomotors. 

Standard RC telemetry is utilized for wireless manual input and real-time monitoring. Ultimately, this integrated setup (detailed in Fig.~\ref{fig:electronic}) ensures reliable propulsion and precise tendon actuation, validating that morphology-driven maneuverability in soft UAVs does not necessitate custom hardware ecosystems.
%
%%%%%%%%%%%%%%%  Experiment  %%%%%%%%%%%%%%%%%%%%%%%%%%
\section{Experiments}\label{sec:MainResults}
%%%%%%%%%%%%%%%% FIGURE 1 %%%%%%%%%%%%%%%%%%%%%
\begin{figure*}[ht]
    \centering
    % --- Subfigura 1 ---
    \begin{subfigure}[t]{0.45\textwidth}
        \centering
        \includegraphics[width=\textwidth]{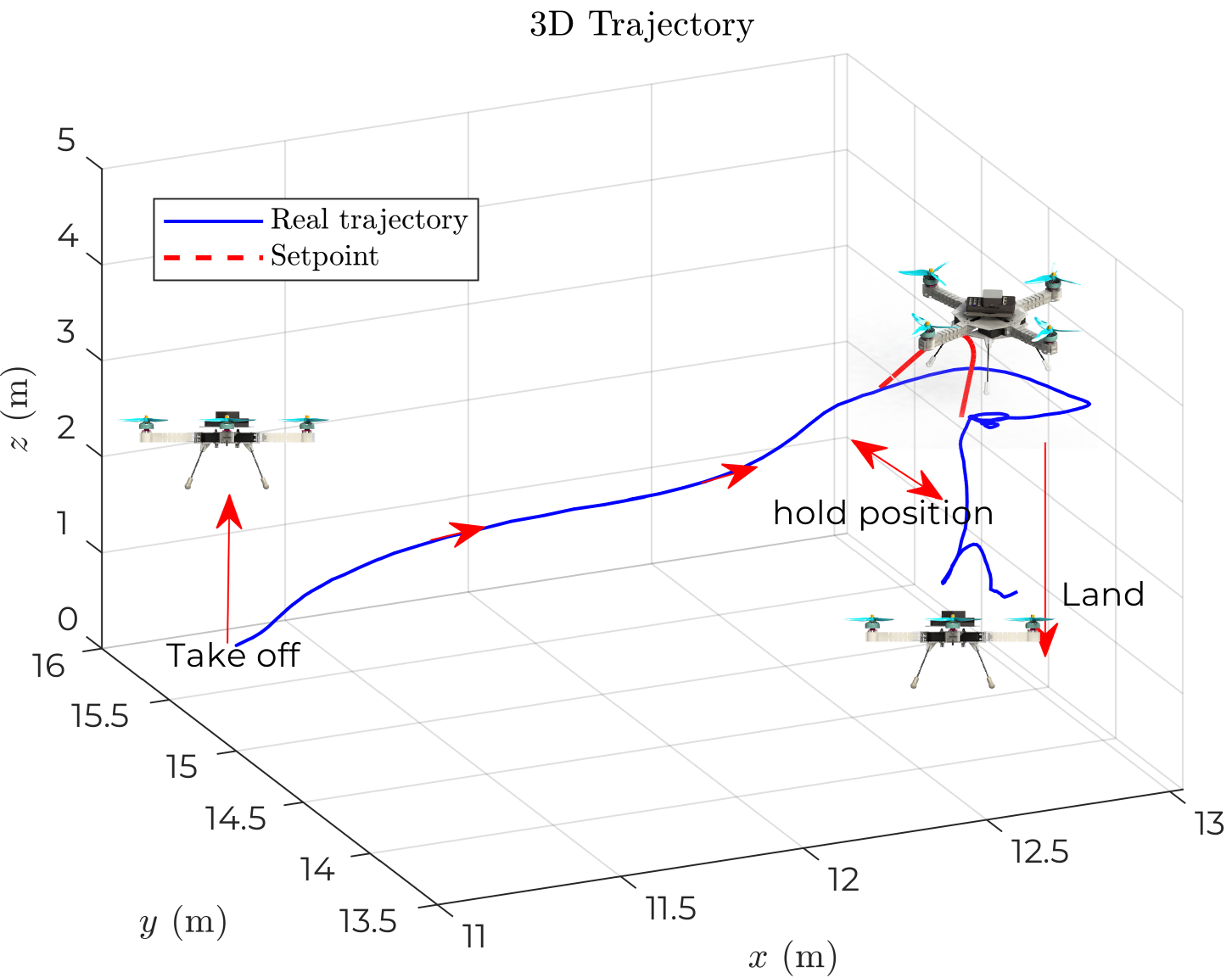}
        \caption{3D trajectory}
        \label{fig:map}
    \end{subfigure}
    \hfill
    % --- Subfigura 2 ---
    \begin{subfigure}[t]{0.50\textwidth}
        \centering
        \includegraphics[width=\textwidth]{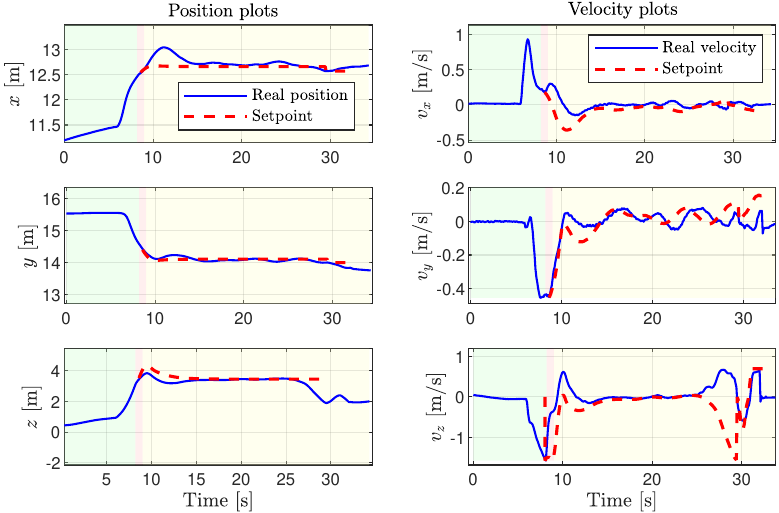}
        \caption{Position and velocity tracking}
        \label{fig:Ztrajectory}
    \end{subfigure}
    \caption{Hover stability experiment with flexible arms. (a) 3D trajectory during take-off, hover, and landing (blue: actual, red dashed: setpoint); Background colors denote the flight modes (green: manual, yellow: hold), and the transition between modes is indicated in red. (b) Position (left) and velocity (right) tracking along $x,y,z$, showing smooth setpoint following across mode transitions.}
    \label{fig:map_bokeh}
\end{figure*}
%%%%%%%%%%%%%%%%%%%%%%%%%%%%%%%%%%%%%%%%%%%%%

To evaluate the \textit{QuadSoft} platform under realistic, unconstrained conditions, we conducted outdoor flight tests subjected to stochastic wind disturbances. Unlike indoor trials that rely on high-precision Motion Capture (MoCap) systems, these outdoor experiments demonstrate the system's operational autonomy using only on-board sensing (IMU/GPS) and a standard, unmodified PX4 control stack. We focused on two critical aspects of soft-rigid hybrid flight:
\begin{itemize}
    \item \textbf{Baseline Hover Stability:} The platform's ability to reject external perturbations and maintain stable setpoints despite the inherent compliance of the soft arms.
    \item \textbf{Morphology-Induced Translation:} The dynamic response of the vehicle when continuous-curvature arm bending is actively used to generate horizontal displacement without altering the vehicle's global pitch/roll attitude.
\end{itemize}

\subsection{Hover Stability}\label{sec:subsectionTakeof}
We first evaluated the QuadSoft’s ability to maintain stable hovering in a zero-curvature configuration. The sequence consisted of a take-off to 4~m, a 15~s hover in \texttt{HOLD} mode, and a manual landing under outdoor conditions with light wind gusts. 

Fig.~\ref{fig:map_bokeh}(a) shows the 3D trajectory, demonstrating that the actual path closely follows the setpoints with negligible lateral drift. Fig.~\ref{fig:map_bokeh}(b) presents the corresponding position and velocity profiles. The controller maintains near-zero steady-state error during the hover phase and recovers seamlessly after mode transitions. 

Crucially, high-speed footage and telemetry confirmed that passive arm deflections at full hovering thrust remained below $1^{\circ}$. This validates the efficacy of the anisotropic structural design (Section~\ref{ssec:StructuralDesign}), proving that the requisite flexibility for tendon actuation does not introduce unmodeled aeroelastic oscillations or degrade the baseline stability expected from a rigid quadrotor.

\subsection{Morphology-Induced Translation}\label{sec:subsectionTraj} 
The core validation of our continuum approach involved generating horizontal displacement purely through tendon-driven arm actuation. Tests were performed in \texttt{ALTITUDE HOLD} mode, isolating horizontal thrust-vectoring from $z$-axis altitude control.

%%%%%%%%%% FIGURE 2 %%%%%%%%%%%%
\begin{figure*}[ht]
    \centering
    % --- Subfigura 1 ---
    \begin{subfigure}[ htb!]{0.50\textwidth}
        \centering
        \includegraphics[width=\textwidth]{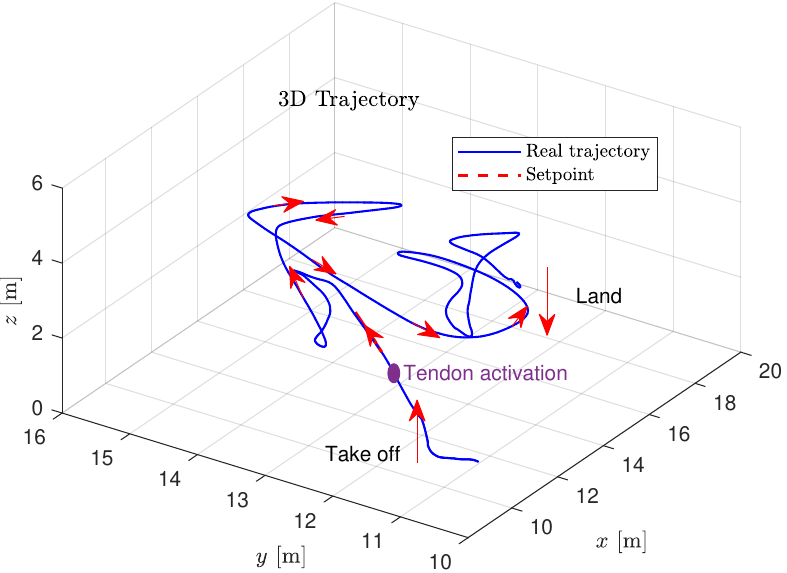}
        \caption{3D trajectory}
        \label{fig:map1}
    \end{subfigure}
    \hfill
    % --- Subfigura 2 ---
    \begin{subfigure}[ htb!]{0.48\textwidth}
        \centering
        \includegraphics[width=\textwidth]{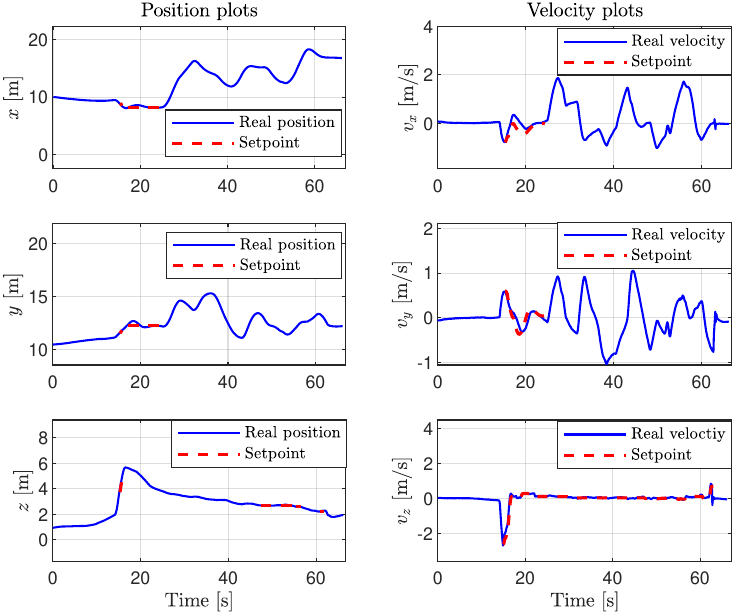}
        \caption{Position and velocity tracking}
        \label{fig:Ztrajectory1}
    \end{subfigure}

    \vspace{0.5cm} % espacio entre filas

    % --- Subfigura 3 ---
    \begin{subfigure}[ htb!]{1\textwidth}
        \centering
        \includegraphics[width=\textwidth]{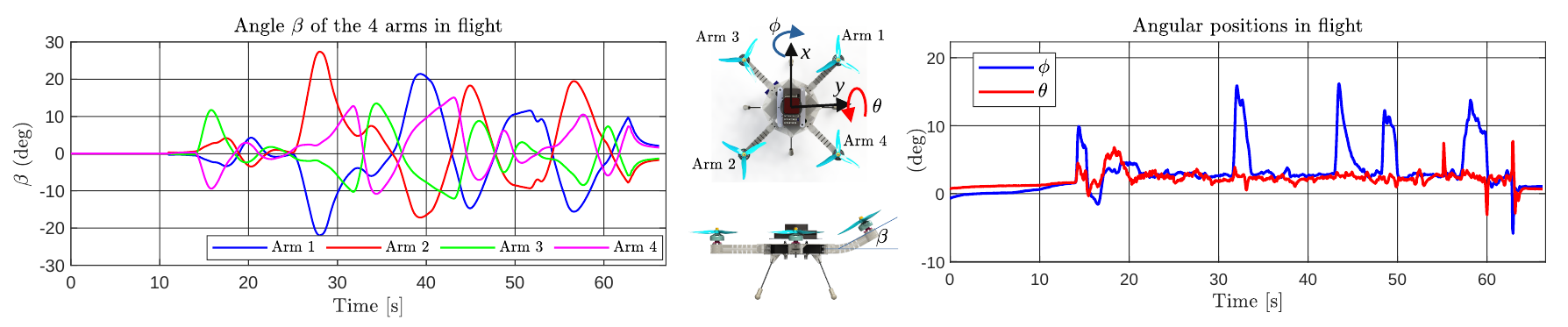}
        \caption{Arm curvatures and pitch/roll angles}
        \label{fig:ang_arms}
    \end{subfigure}
    \caption{Morphology-induced translation with tendon activation. (a) 3D trajectory (arrows indicate motion direction; purple dot marks tendon activation). (b) Position and velocity tracking showing steady altitude during translation. (c) Arm curvature angles $\beta_i$ and vehicle attitude ($\phi$, $\theta$), confirming the link between continuum actuation and horizontal displacement with minimal attitude coupling.}
    \label{fig:map_bokeh1}
\end{figure*}
%%%%%%%%%%%%%%%%%%%%%%%%%%%%%%%%

As illustrated in Fig.~\ref{fig:map_bokeh1}(a) and (b), upon tendon activation, the vehicle achieved smooth $x$–$y$ translation while strictly maintaining its target altitude. A gradual velocity increase coincided precisely with the onset of arm curvature. 

Fig.~\ref{fig:map_bokeh1}(c) presents the transient response during reconfiguration, detailing the measured arm curvatures ($\beta_i$) alongside the vehicle's attitude ($\phi$, $\theta$). As the arms actively bend to induce forward motion, the global pitch and roll remain remarkably stable, staying within a $\pm5^{\circ}$ envelope. The brief excursions up to $10^{\circ}$ correlate with corrective actions against wind gusts rather than actuation-induced instability. 

Most importantly, the data demonstrates that during the dynamic transition phase (from $0^\circ$ to the validated $22^\circ$ curvature), the change in the vehicle's thrust vector is smoothly compensated by the native PX4 rate controllers. This confirms that the structural damping of the soft arms prevents resonance modes, allowing morphology-driven directional motion with minimal attitude deviations. Ultimately, these unconstrained outdoor trials establish a robust physical baseline, proving the viability of tendon-driven continuum morphing for adaptive aerial systems. 
\section{Conclusion}\label{conclusion}
The experimental evaluation of the \textit{QuadSoft} platform leads to several critical conclusions regarding the viability and control of soft aerial vehicles. 

First, the outdoor flight data demonstrates that the inherent vulnerabilities of soft structures—namely, aeroelastic vibrations and uncontrolled deformations—can be successfully mitigated through anisotropic structural design. Because the vehicle maintained stable hovering and its global attitude deviations remained within $\pm5^{\circ}$ during dynamic reconfiguration, we conclude that the TPU-carbon hybrid arms act as an effective vibration damper. 

Second, the successful morphology-induced translation confirms that a \textbf{constant-curvature continuum approach} is a highly effective mechanism for thrust vectoring. By generating forward and lateral displacements while maintaining a near-level pitch and roll, the results establish that tendon-actuated soft arms can practically decouple a vehicle's position from its orientation. 

Finally, achieving these stable transitions under stochastic wind disturbances using a standard, unmodified PX4 autopilot leads to a fundamental conclusion: morphological compliance can be seamlessly integrated into existing flight stacks as a physical asset rather than a control disturbance. Ultimately, this work provides a validated physical baseline for fully actuated soft UAVs. Based on these findings, future research will focus on developing custom control allocation matrices (mixers) within the open-source PX4 ecosystem to fully exploit the platform's 6-DOF maneuverability.
%%%%%%%%%%%%%%%%%%% FOTO DEL EXPERIMENTO %%%%%%%%%%%%%
\begin{figure*}[ht]
  \centering
  \includegraphics[width=\textwidth]{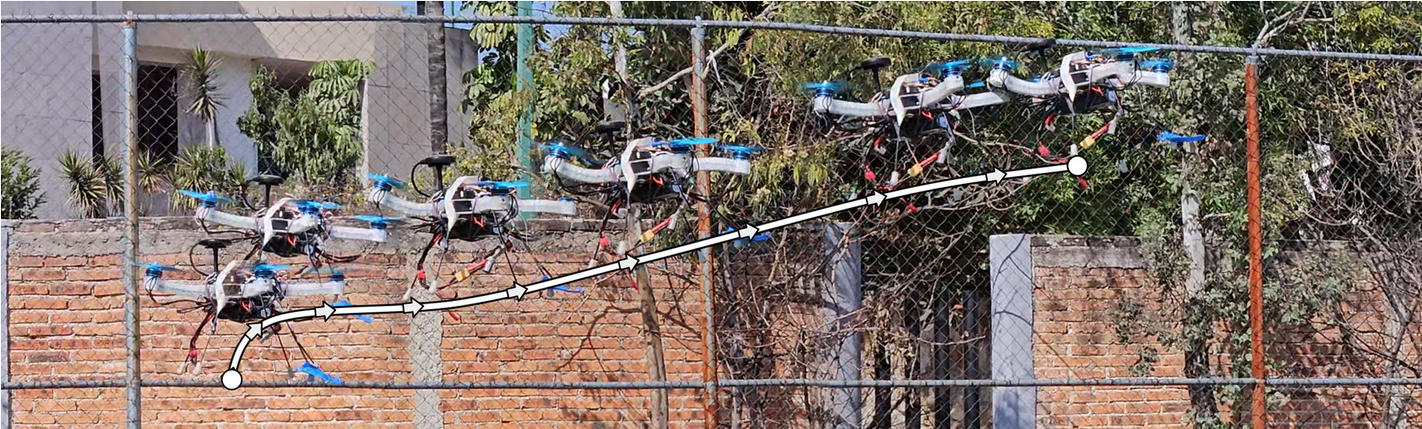}
  \caption{Sequence of an outdoor flight demonstrating forward displacement generated by tendon-driven arm actuation under altitude hold. The snapshots illustrate the QuadSoft’s trajectory during morphological reconfiguration, highlighting controlled horizontal motion achieved without affecting altitude.}
  \label{fig:hold}
\end{figure*}
%%%%%%%%%%%%%%%%%%% FOTO DEL EXPERIMENTO %%%%%%%%%%%%%

%
%\subsection{Future Work}
%Future efforts will follow three main directions. First, we will develop nonlinear and morphology-aware control algorithms that explicitly account for the coupling between arm curvature and vehicle dynamics. This includes control allocation strategies that translate tendon-driven deformations directly into thrust and moment generation, enabling precise and adaptive flight. Second, we plan to embed sensing into the soft arms—such as curvature, strain, and contact-force sensors—to provide feedback for real-time adaptation, improving resilience and ensuring safer operation in unstructured or dynamic environments. Third, we will extend testing to more demanding and application-oriented scenarios, including aerial manipulation, perching on irregular structures, and payload delivery. These experiments will demonstrate how tendon-driven morphologies can not only match but potentially surpass rigid quadrotors in tasks that require adaptability, robustness, and precision. In the longer term, such advances may open a pathway toward a new class of aerial robots that are intrinsically reconfigurable and capable of interacting seamlessly with complex environments.
%
%\textcolor{red}{Debe haber al menos 30 referencias \textbf{de soft robotics} en el estado del arte.}
%
\bibliographystyle{IEEEtran}
\bibliography{IEEE_biblioCONVERTIBLE}

\vfill

\end{document}